\documentclass[sigconf]{acmart}

\usepackage{booktabs} % For formal tables
\usepackage{amsmath,amssymb,amsfonts}
\usepackage{algorithmic}
\usepackage{graphicx}
\usepackage{textcomp}
\usepackage{epsfig,subfig}		% Inclusão de figuras
\usepackage{tabularx} 					% Para poder ter tabelas com colunas de largura auto-ajustável
\usepackage{multirow}					% Pacote para multiplas linhas em tabelas
\usepackage{scalefnt}
\usepackage{adjustbox}
\usepackage{threeparttable}
\usepackage[T1]{fontenc}
\usepackage[brazil]{babel} 
\usepackage[utf8]{inputenc}
\usepackage{xcolor,colortbl}

\usepackage{xcolor}
\usepackage[textsize=tiny]{todonotes}

\setcopyright{acmcopyright}
\acmConference[SBQS'19]{Simp{\'o}sio Brasileiro de Qualidade de Software}{Outubro 2019}{Fortaleza, Cear{\'a}, Brasil}
\acmYear{2019}
\copyrightyear{2019}

\acmPrice{15.00}

\copyrightyear{2019} 
\acmYear{2019} 
\acmConference[SBQS'19]{XVIII Brazilian Symposium on Software Quality}{October 28-November 1, 2019}{Fortaleza, Brazil}
\acmBooktitle{XVIII Brazilian Symposium on Software Quality (SBQS'19), October 28-November 1, 2019, Fortaleza, Brazil}
\acmPrice{15.00}
\acmDOI{10.1145/3364641.3364647}
\acmISBN{978-1-4503-7282-4/19/10}

\begin{document}

\title{Effects of Visualizing Technical Debts on a Software \\Maintenance Project}

\author{Ronivon Silva Dias}
\affiliation{
	\institution{Universidade Federal do Piauí} 
	\city{Teresina}
	\state{Piauí}
	\country{Brasil}
}
\email{ronivon@ufpi.edu.br}

\author{Pedro de Alc\^{a}ntara dos Santos Neto}
\affiliation{
	\institution{Universidade Federal do Piauí} 
	\city{Teresina}
	\state{Piauí}
	\country{Brasil}
}
\email{pasn@ufpi.edu.br}

\author{Irvayne Matheus de Sousa Ibiapina}
\affiliation{
	\institution{Universidade Federal do Piauí} 
	\city{Teresina}
	\state{Piauí}
	\country{Brasil}
}
\email{irvaynematheus@gmail.com}

\author{Guilherme Amaral Avelino}
\affiliation{
	\institution{Universidade Federal do Piauí} 
	\city{Teresina}
	\state{Piauí}
	\country{Brasil}
}
\email{gaa@ufpi.edu.br}

\author{Otávio Cury da Costa Castro}
\affiliation{
\institution{Universidade Federal do Piauí} 
	\city{Teresina}
	\state{Piauí}
	\country{Brasil}
}
\email{otaviocury.oc@gmail.com}

\renewcommand{\shortauthors}{D. Ronivon et al.}

\begin{abstract}
The technical debt (TD) metaphor is widely used to encapsulate numerous software quality problems. She describes the trade-off between the short term benefit of taking a shortcut during the design or implementation phase of a software product (for example,  in order to meet a deadline) and the long term consequences of taking said shortcut, which may affect the quality of the software product. TDs must be managed to guarantee the software quality and also reduce its maintenance and evolution costs. However, the tools for TD detection usually provide results only considering the files perspective (class and methods), that is not usual during the project management. In this work, a technique is proposed to identify/visualize TD on a new perspective: software features. The proposed technique adopts Mining Software Repository (MRS) tools to identify the software features and after the technical debts that affect these features. Additionally, we also proposed an approach to support maintenance tasks guided by TD visualization at the feature level aiming to evaluate its applicability on real software projects. The results indicate that the approach can be useful to decrease the existent TDs, as well as avoid the introduction of new TDs.
\end{abstract}




\maketitle

%!TEX root = ../SBQS_2019_DT.tex

\section{Introdução}
\label{sec_introducao}

As equipes de desenvolvimento de \textit{software} enfrentam com frequência o desafio de entregar produtos de \textit{software} em prazos apertados, enquanto tentam manter um padrão de qualidade. Como resultado, surgem artefatos de baixa qualidade, pois  os desenvolvedores tendem a  deixarem de lado as boas práticas de engenharia de \textit{software} e princípios de programação para entregar o produto dentro do prazo estipulado, ainda que imaturo sob diversos aspectos ~\cite{brown,zazworka}. Esse processo pode muitas vezes resultar na introdução do que é conhecido como Dívida Técnica (DT) ~\cite{Cunningham:1992:WPM:157709.157715}, que ocasiona problemas para o \textit{software} em futuras manutenções e consequentemente, compromete a sua evolução~\cite{zazworka}.

%A DT consiste em um conjunto de pendências ocorridas durante o desenvolvimento de \textit{software}, proporcionado por uma escolha que é interessante a curto prazo, levando-se em consideração a produtividade, tempo de entrega, redução de esforço e custo, mas que aumenta a complexidade e é mais oneroso a longo prazo ~\cite{kruchten2012technical}.  Além disso, os efeitos negativos da DT na evolução do \textit{software} podem comprometer a qualidade e reduzir a manutenibilidade do código, e ainda contribuir para um ambiente favorável ao surgimento de defeitos ~\cite{Cunningham:1992:WPM:157709.157715}, gerando problemas para o ecossistema do desenvolvimento ~\cite{tom}. Por conta disso, é crescente a atenção para o gerenciamento de DTs, tanto na comunidade de desenvolvimento de \textit{software}, quanto na científica ~\cite{brown}.

Para lidar com a dívida existente em um sistema de \textit{software} ou para evitar que dívidas potenciais sejam introduzidas, Li et al. ~\cite{Zengyang:2015} identificaram e propuseram oito atividades de manejo de DT. Essas oito atividades são: a identificação, medição, priorização, prevenção, monitoramento, reembolso, documentação e comunicação da DT. Dentre essas atividades, a identificação (detecção de DT usando técnicas como análise de código estático), medição (quantificação de DT usando técnicas de estimativa) e reembolso (resolução de DT por técnicas como reengenharia ou refatoração) recebem maior atenção da comunidade técnica e científica, com apoio de ferramentas e abordagens apropriadas~\cite{Zengyang:2015}. 

No entanto, as ferramentas para detecção de DT geralmente fornecem resultados só considerando a perspectiva de arquivos de código fonte (classes e métodos) ~\cite{SonarQube,marple,anacodedebt,tedma,mendesvisminertd,Debtflag}, que por si só, tende a não ser apropriada considerando um contexto real. É comum que durante a manutenção de \textit{software}, não se analise um arquivo de código fonte em específico, mas sim uma funcionalidade, que pode ser composta por diversos arquivos de código fonte, em diversas pastas diferentes. Ou seja, a perspectiva de funcionalidade é a mais comumente ligada à manutenção, especialmente no que se refere à distribuição de tarefas~\cite{Kersten2005}.

Na prática, quando um desenvolvedor recebe uma tarefa que reporta um problema ou uma recomendação de melhoria, é comum que ela cite a funcionalidade associada, sem qualquer ligação com os arquivos de código fonte ligados ao caso. Por conta disso, faz-se necessária uma nova perspectiva sobre a gestão de DT, visando facilitar as decisões gerenciais associadas.

Wohlin et al.~\cite{Wohlin} afirmam que a única avaliação real de um processo/método é ter pessoas usando-o, já que o mesmo é apenas uma descrição até que as pessoas o utilizem. De fato, o caminho da subjetividade para a objetividade é pavimentado por testes ou comparação empírica com a realidade ~\cite{Juzgado}. Neste sentido, foi proposto um método que visa identificar e visualizar as DT em nível de funcionalidade e uma abordagem que faz uso desse método, para que assim pudéssemos testá-lo em um ambiente real de desenvolvimento.

Este trabalho tem como objetivo apoiar a gestão de DT, por meio da exposição de tais dívidas associadas a tarefas de manutenção e evolução de \textit{software}, na tentativa de influenciar a eliminação dessas dívidas. Destacam-se como principais contribuições deste trabalho as seguintes:

\begin{itemize}
    \item Definição de um método para identificar as dívidas técnicas utilizando a perspectiva de funcionalidades;
    \item Extensão da CoDiVision para implementar o método de visualização de dívida técnica por funcionalidades;
    \item Criação de uma abordagem para manutenção/evolução de \textit{software} com foco na redução de DT;
    \item Realização de um estudo de caso avaliando a abordagem proposta neste trabalho em um cenário industrial.
\end{itemize}

Este trabalho está organizado da seguinte forma: a Seção ~\ref{sec_referencial_teorico} apresenta os principais conceitos relacionados ao trabalho; a Seção ~\ref{sec_relacionados} apresenta alguns estudos relacionados ao tema; na Seção ~\ref{sec:metodo} é apresentado o método proposto; que na Seção ~\ref{sec:tool} é apresentado o modulo TDVision que é um extensão da CoDiVision para implementar o método proposto; na Seção ~\ref{sec:abordagem} é apresentadoa a abordagem criada com foco na redução de DT; na Seção ~\ref{sec:avaliacao} é descrito um estudo de caso; na Seção ~\ref{sec:ameacas} são apresentadas as ameaças a validade; e, por fim, a Seção ~\ref{sec:conclusao} apresenta as conclusões e direções para trabalhos futuros.

%!TEX root = ../SBQS_2019_DT.tex

\section{Referencial Teórico}
\label{sec_referencial_teorico}

Nesta seção serão descritos alguns conceitos importantes para o entendimento do método proposto.

\subsection{Dívida Técnica}

Dívida Técnica representa uma metáfora que se refere às consequências do desenvolvimento de \textit{software} deficiente, imaturo ou impróprio ~\cite{Cunningham:1992:WPM:157709.157715}. Alguns autores caracterizam as DT como sendo qualquer parte do \textit{software} atual que é considerado abaixo do um bom nível sob uma perspectiva técnica ~\cite{tom}.

As DT podem ser classificadas, de acordo com a natureza da sua concepção, em 2 grupos: intencional e não intencional ~\cite{mcconnell2007technical}. A DT causada de forma intencional normalmente ocorre por decisões conscientes, feita pela equipe, com o objetivo de otimizar o presente em detrimento ao futuro. Já a DT classificada como não intencional normalmente ocorre por conta de trabalhos de baixa qualidade ou carência técnica de membros da equipe de desenvolvimento. 

Alves et al.~\cite{alves2016identification} sintetizam algumas dimensões da Dívida Técnica, categorizando-as em 15 tipos: Arquitetura, Construção, Projeto, Código, Defeito, Teste, Documentação, Infraestrutura, Pessoas, Processos, Requisitos, Serviço, Automação de Teste, Usabilidade e Versionamento. Essas DT podem ser caracterizados por indicadores identificados através de cálculo de métricas de código, detecção de \textit{code smells} e/ou análise de comentários.

Dentre as dívidas que podem surgir no código fonte de um projeto, podemos destacar a Dívida de Código. Neste trabalho, a caracterização de Dívidas de Código e Design é baseado no conjunto de 18 métricas de código e 7 \textit{code smells}~\cite{170925}. Na Tabela ~\ref{tabela_smells} são apresentados os \textit{code smells} considerados neste trabalho e suas respectivas descrições.

\begin{table}[h]
	\centering
	\caption{Tipos de code smells considerados neste trabalho.}
	\label{tabela_smells}
	\resizebox{.4\textwidth}{!}{
		\begin{tabular}{|c|c|}
			\hline
			\textbf{Code Smell} & \textbf{Descrição} \\ \hline
			\textit{\textbf{God Class}} & \begin{tabular}[c]{@{}c@{}}Caracterizado por classes que centralizam \\ a inteligência de um sistema.\end{tabular} \\ \hline
			\textit{\textbf{Brain Method}} & \begin{tabular}[c]{@{}c@{}}Caracterizado por métodos que centralizam \\ as funcionalidades de uma classe.\end{tabular} \\ \hline
			\textit{\textbf{Brain Class}} & \begin{tabular}[c]{@{}c@{}}Caracterizado por classes que acumulam \\ uma excessiva quantidade de inteligência, \\ normalmente possuindo muitos Brain Methods.\end{tabular} \\ \hline
			\textit{\textbf{Data Class}} & \begin{tabular}[c]{@{}c@{}}Caracterizado por classes que somente \\ armazenam dados e não possuem \\ funcionalidades complexas. \\ Elas oferecem mais dados do que serviços.\end{tabular} \\ \hline
			\textit{\textbf{Conditional Complexity}} & \begin{tabular}[c]{@{}c@{}}Caracterizados por métodos que possuem \\ muitas estruturas condicionais.\end{tabular} \\ \hline
			\textit{\textbf{Long Method}} & \begin{tabular}[c]{@{}c@{}}Caracterizado por métodos que possuem \\ muitas linhas de código.\end{tabular} \\ \hline
			\textit{\textbf{Feature Envy}} & \begin{tabular}[c]{@{}c@{}}Caracterizado por métodos que acessam \\ muitos dados de outras classes e poucos da sua própria classe.\end{tabular} \\ \hline
		\end{tabular}
	}
\end{table}

\subsection{Mineração de Repositório de \textit{software}}
\label{cap:fundamentacaoteorica:sec:scv}

Alguns estudos ~\cite{PRESSMAN,SPINELLIS} afirmam que projetos de \textit{software}, especialmente aqueles com equipes de grande porte, que não utilizam o Sistema de Controle de Versão (SCV), muitas vezes encontram problemas ao longo do tempo, como artefatos inconsistentes. O SCV, além de fornecer suporte ao desenvolvimento simultâneo por um grupo de desenvolvedores, permite o acompanhamento de tal desenvolvimento, fornecendo um histórico de todas as mudanças junto aos desenvolvedores que os criaram.

Com a adoção de um SCV, o repositório de \textit{software} se torna uma fonte significativa de informações, incluindo todas as versões de arquivos, as linhas de código que foram alteradas em cada versão e informações sobre quando as alterações foram realizadas e por quem~\cite{SPINELLIS}. Para Storey et al. ~\cite{STOREY}, o repositório é uma memória organizacional que pode ser acessada para avaliar a evolução do \textit{software}.

A Mineração de Repositórios de \textit{software} (MRS) é um campo de pesquisa que analisa dados históricos provenientes de repositórios de \textit{software}, tais como Sistemas de Controle de Versão (SCV) e ferramentas para gestão de tarefas. As pesquisas nessa área têm como objetivo transformar os repositórios em ativos para apoiar nas atividades inerentes ao processo de desenvolvimento. A partir de dados provenientes de tais repositórios é possível descobrir padrões e informações úteis para a compreensão do \textit{software} e do ecossistema envolvido na sua construção ~\cite{hassan}.

As etapas típicas de um processo de análise de dados a partir da MRS são: i) Extração; ii) Modelagem de Dados; iii) Síntese; e iv) Análise ~\cite{hemmati2013msr}. Na primeira etapa é realizada a extração de dados “brutos” de vários tipos de repositórios. Na etapa de Modelagem é realizada uma preparação dos dados para serem utilizados. Na etapa de Síntese é realizado um processamento dos dados extraídos. Finalmente, é necessária a etapa de Análise e interpretação dos resultados para formulação de conclusões. %Na Figura ~\ref{fig_mrs} são apresentadas as etapas citadas.

%\begin{figure}[h]
%	\begin{center}
%		\includegraphics[scale=0.48]{Figuras/mrs}
%		\caption{Etapas típicas em um processo de MRS.}
%		\label{fig_mrs}
%	\end{center}
%\end{figure}

O processo de Mineração de Dados pode ser conduzido de várias maneiras distintas, de acordo com os objetivos que se deseja alcançar com a aplicação da técnica. Tais objetivos podem ser organizados em dois grandes grupos~\cite{ZAKI:2003}:
\begin{itemize}
\item[i.]Atingir uma capacidade preditiva confiável, ou seja, buscar responder quais fenômenos podem vir a acontecer; e

\item[ii.]Alcançar uma descrição compreensível, ou seja, identificar a razão de fenômenos já conhecidos acontecerem da forma como acontecem;

\end{itemize}

Pesquisadores no campo da engenharia de \textit{software} têm desenvolvido abordagens para extrair informações pertinentes e descobrir relações e tendências de repositórios no contexto da evolução do \textit{software}~\cite{KAGDI:2007}.  Hassan e Xie ~\cite{hassan} analisaram a evolução das pesquisas sobre a mineração de repositório de \textit{software}, frisando a importância das informações que estão em repositórios de \textit{software} que são extremante úteis para tomada de decisões em projetos de \textit{software}. A mineração de repositório de \textit{software} permite a utilização de indicadores mais precisos do que simplesmente a intuição e a experiência de desenvolvedores e gestores nas tomadas de decisão. Eles apontaram que ferramentas online e ou integradas ao ambiente de desenvolvimento de \textit{software} estão surgindo. Ou seja, novas oportunidades estão sendo criadas para serem exploradas nesta área.

%!TEX root = ../SBQS_2019_DT.tex

\section{Trabalhos Relacionados}
\label{sec_relacionados}

Um dos objetivos deste trabalho é propor um método para identificação e visualização de DT sobre uma nova perspectiva: funcionalidade de \textit{software}. Essa nova visão de DT busca facilitar as decisões sobre seu gerenciamento. Nas pesquisas realizadas na literatura não foram identificados trabalhos que utilizam a visão de funcionalidade para atingir esse objetivo. Por conta disso, é apresentado a seguir, alguns trabalhos e ferramentas que foram consideradas na elaboração da nossa abordagem, mas que não possuem o mesmo objetivo do trabalho aqui proposto.

%Palomba e demais autores ~\cite{Palomba:2014} investigaram quais \textit{Code Smells} são considerados pelos desenvolvedores como os mais prejudiciais. Os desenvolvedores receberam trechos de código de três sistemas com doze tipos de \textit{code smells} e foram solicitados a classificar a severidade dos \textit{smells}s. Os resultados sugerem que existem algumas categorias de \textit{code smells} que: i) são altamente percebidos e identificados pelos desenvolvedores; ii) criam mais preocupações para os desenvolvedores com mais experiência e conhecimento do sistema; e iii) são classificados com valores de severidade altos. Os \textit{Code Smells} com tais propriedades são \textit{complex class}, \textit{god class}, \textit{long method} e \textit{spaghetti code}. Nesse mesmo estudo foi constatado que existe um grupo de \textit{Code Smells} para os quais a percepção de severidade varia caso a caso. Esses \textit{smells} são \textit{feature envy}, \textit{refused bequest} e \textit{speculative generality}.

O \textit{SonarQube}\footnote{https://docs.sonarqube.org/} é uma das ferramentas mais conhecidas e utilizadas na indústria de \textit{software} para realizar inspeção de qualidade de código ~\cite{campbell2013sonarqube}. É uma plataforma aberta para gerenciamento da qualidade do código. A qualidade do \textit{software} é avaliada de acordo com sete eixos: comentários, potenciais erros, complexidade, testes de unidade, duplicações, regras de codificação, arquitetura e projeto ~\cite{SonarQube}. 

A plataforma Sonar utiliza o \textit{Technical Debt Plugin} para o cálculo da DT do projeto. Primeiramente, são calculados os valores de seis eixos básicos: duplicação, violação, complexidade, cobertura, documentação  e projeto. Em seguida, essas métricas são somadas para fornecer uma métrica global. Os detalhes da composição dessa métrica são descritos abaixo: 

\begin{itemize}
\item Proporção da dívida: fornece um percentual da DT atual em relação ao total da dívida possível para o projeto. 
\item O custo para pagar as DT: determina o custo, em valor monetário, necessário para limpar todos os defeitos em cada eixo. 
\item O trabalho para pagar as DTs: fornece o esforço para pagar as DTs expresso utilizando a métrica homens/dia. 
\item A repartição: apresenta um gráfico com a distribuição do débito nos seis eixos de qualidade. 
\end{itemize}

Assim como o \textit{SonarQube}, existem diversas ferramentas que visam de forma automatizada auxiliar nas atividades de gestão da dívida técnica. Essas ferramentas exploram diferentes técnicas para detectar DT: algumas são baseadas em métricas (\textit{CodeVizard}\footnote{http://www.nicozazworka.com/tool-development/}, \textit{MARPLE}~\cite{marple}, \textit{AnaConDebt}~\cite{anacodedebt}, \textit{TEDMA} ~\cite{tedma}, \textit{VisminerTD} ~\cite{mendesvisminertd}), enquanto outras são plugins integrados a uma ferramenta do ambiente de desenvolvimento (\textit{FindBugs}\footnote{http://findbugs.sourceforge. net/}, \textit{DebtFlag} ~\cite{Debtflag}) outras usam análise estática do código para identificar oportunidades de refatoração (\textit{JDeodorant}\footnote{https://marketplace.eclipse.org/content/jdeodorant}, \textit{AnaConDebt}~\cite{anacodedebt}, \textit{CodeScene}\footnote{https://codescene.io/}. Porém todas elas identificam e priorizam as DT em nível de arquivo (classe) o que pode tornar difícil à negociação com os stakeholders  sobre o pagamento das mesmas.

%Brown et al.~\cite{brown} realizaram um estudo a respeito de uma modelagem para pagamento da DT baseada na análise de impactos em longo prazo. Os resultados apontam que há divergência quanto às intenções de investimento de qualidade por parte de desenvolvedores, engenheiros e gerentes, que acabam não investindo em gerência da DT por não identificarem resultados em curto prazo.

Klinger e demais autores~\cite{KLINGER} realizaram um estudo sobre a premissa de que dívida técnica deve ser usada como uma ferramenta para conhecimento de benefícios não descobertos e que surge das más escolhas de desenvolvimento por parte de arquitetos e desenvolvedores. Foram feitas entrevistas com 4 arquitetos de \textit{software} da IBM com diferentes níveis de experiência, com o objetivo de obter respostas sobre como os stakeholders influenciaram na dívida, e quais as estratégias para gerencia-la e quantificá-la. Juntando-se as respostas dos entrevistados, conclui-se que: i) calcular a dívida, quando induzida por questões gerenciais é um desafio, havendo necessidade, em alguns casos, de se adquiri-la em razão de aceitação e sucesso; ii) as decisões são tomadas de maneira \textit{ad-hoc}, ou seja, não há processos formais de notificar a dívida, tornando-a mais invisível ainda; iii) há falhas em comunicar a \textit{stakeholders} a dívida, bem como lhes faltam compreensão, pois os mesmos priorizam funcionalidade à qualidade.

Nosso método se difere das demais ferramentas por propor a identificação de dívidas técnicas por funcionalidade e não por arquivo. Esse método tem como objetivo auxiliar na comunicação das dívidas aos envolvidos no processo.

%!TEX root = ../SBQS_2019_DT.tex

\section{Método  Proposto}
\label{sec:metodo}

Dado um projeto que contém DT, o método proposto neste trabalho visa identificar e visualizar as mesmas em nível de funcionalidade. 

Para isso, o método proposto possui 3 etapas, sendo inicialmente feita a extração de dados a partir de um sistema de controle de versão. Em seguida, são identificadas as funcionalidades que pertencem a um sistema de \textit{software}. Com as funcionalidades e suas respectivas classes identificadas, é realizada a etapa de cálculo  do total de dívidas técnicas presentes nas funcionalidades e gerado uma lista ordenada. Na Figura ~\ref{fig_abordagem} são apresentadas as etapas do método.

\renewcommand{\thefigure}{\arabic{figure}}
\begin{figure}[h]
	\begin{center}
		\includegraphics[scale=0.7]{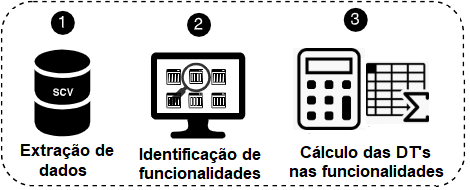}
		\caption{Etapas do método proposto.}
		\label{fig_abordagem}
	\end{center}
\end{figure}

Nas próximas subseções serão descritos em mais detalhes cada uma das etapas do método proposto.

\subsection{1ª Etapa - Extração de Dados}
\label{sec:etapa1_extracao_dados}

O objetivo desta etapa é realizar a extração de dados a partir de um Sistema de Controle de Versão (SCV). Os arquivos podem ser obtidos dos principais tipos de sistemas de versionamento: Git ou SVN. Os dados extraídos são: a) Histórico de \textit{commits} realizados; b) \textit{Diffs} (apresenta a diferença entre duas versões de um mesmo arquivo) de arquivos alterados e c) Arquivos de código fonte do sistema. Para a realização desta etapa de extração de dados, são utilizados os módulos já existentes na ferramenta \textit{CoDiVision}~\cite{codivision}. A ferramenta utiliza-se das API's JGit\footnote{\url{http://www.eclipse.org/jgit/}.} e SVNKit\footnote{\url{https://svnkit.com/}.} para extração de repositórios Git e SVN, respectivamente.

\subsection{2ª Etapa  - Identificação de Funcionalidades}
\label{sec:etapa1_calculo_metricas}

\begin{figure}[htpb]
   \centering
   \includegraphics[scale=.52]{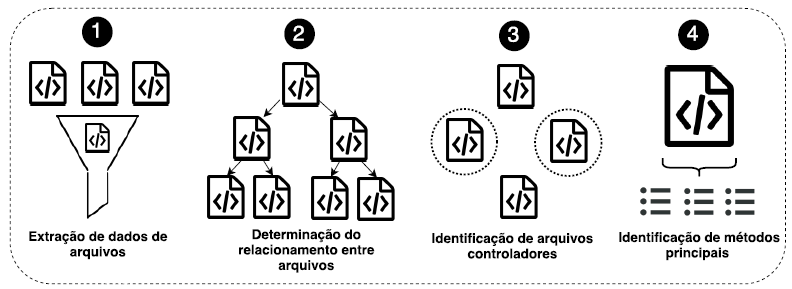}
   \caption{Etapas do processo para identificação de funcionalidades proposto por Vanderson et al.}
   \label{fig:etapas_funcionalidade}
\end{figure}

Nesta etapa serão identificadas as funcionalidades presentes no \textit{software} a ser analisado. Será utilizada a abordagem proposta por Vanderson et al. ~\cite{vanderson:2018}, que está implementada na ferramenta \textit{CoDiVision}~\cite{codivision}, conforme é ilustrado na Figura ~\ref{fig:etapas_funcionalidade} e descrito brevemente a seguir.

\begin{itemize}

  \item[1.] \textbf{Extração de informações dos arquivos de código}: são extraídos informações a partir de arquivos de código fonte Java que foram extraidos na etapa ~\ref{sec:etapa1_extracao_dados}. Isso é feito processando cada classe por um \textit{parser} desenvolvido por meio da API Eclipse JDT. Esse \textit{parser} mapeia o código fonte Java em uma \textit{Abstract Syntax Tree} (AST). As informações são referentes a todos os dados contidos em um arquivo de código, como por exemplo, variáveis e métodos (incluindo seus parâmetros e tipo de retorno). 
  
  \item[2.] \textbf{Determinação do relacionamento entre arquivos}: utilizando as informações extraídas durante a subetapa anterior é identificado o relacionamento entre arquivos, isto é, determina-se para cada arquivo $a$, o conjunto de outros arquivos do sistema referenciados por $a$.

  \item[3.] \textbf{Identificação de arquivos controladores}: identifica arquivos ou classes principais do sistema, denominados aqui de controladores, isto é, aqueles responsáveis por iniciar a execução de funcionalidades do sistema. Para isso, é criado um grafo de relacionamento entre arquivos para então identificar, com base em referências realizadas, quais são arquivos controladores. 
  
  \item[4.] \textbf{Identificação de métodos principais em arquivos controladores}: devem ser identificados os métodos ou funções principais de cada arquivo controlador. Os métodos ou funções principais de um arquivo controlador
  representam os pontos de início de execução das funcionalidades do sistema. Um método principal é identificado por meio de sua acessibilidade (modificador de acesso), número de chamadas à outros métodos e chamadas (invocações) recebidas por outros métodos do mesmo arquivo.

\end{itemize}

Como resultado destas 4 subetapas para identificação de funcionalidades de \textit{software}, é gerado um conjunto definido por $F = \{f_1, f_2, ..., f_n\}$, onde $n$ é o número de funcionalidades identificadas e $f_i$ representa a $i$-ésima funcionalidade. Para cada funcionalidade existe uma lista de arquivos (classes) que a integram.

\subsection{3ª Etapa - cálculo do total de dívidas técnicas}

Com as funcionalidades e os arquivos que as compõem devidamente identificadas, é necessário a identificação e contabilização das dívidas técnicas presentes nas mesmas. Para isso, recomenda-se o uso de uma ferramenta para identificação de DT em cada classe relacionada a uma funcionalidade, para então exibir as DT agrupadas por funcionalidade. Neste trabalho foi utilizada a \textit{API RepositoryMiner}~\cite{repositoryMiner}, que foi integrada a ferramenta \textit{CoDiVision}~\cite{codivision}.

A \textit{RepositoryMiner} é capaz identificar code smells de maneira rápida, automatizada, e com precisão, além de abranger os code smells bem-aceitos na literatura, no qual foram propostos por trabalhos relevantes ~\cite{lanza2007object,KERIEVSKY,FOWLER}, para caracterizar a presença de DT em projetos de software. Ela detecta 7 tipos de \textit{code smells} em projetos desenvolvidos na linguagem Java. Na Tabela~\ref{tabela_smells} pode-se ver quais os \textit{code smells} disponíveis e suas respectivas descrições. 

%Esses \textit{code smells} são calculadas a partir de métricas de \textit{software} extraídos do código fonte do projeto e identificadas de forma estática, sem a necessidade de executar o código.

Na Tabela~\ref{tabela_arquivo} coluna \textbf{A} podemos observar os arquivos da funcionalidade \textbf{Matricula Componente} que foram identificados na etapa ~\ref{sec:etapa1_calculo_metricas} e na coluna \textbf{B}  o total de DT de cada arquivo da funcionalidade que foram encontrados usando a \textit{RepositoryMiner}. Com base nesses dados, podemos realizar o somatório das DT de todos os arquivos e teremos o total de DT da funcionalidade, que nesse exemplo é 11 (ou seja, a funcionalidade \textbf{Matricula Componente} possui 11 DT). Essa mesma operação é feita para todas as funcionalidades e seus arquivos.

\begin{table}[h]
	\centering
	\caption{Cálculo do total de dívidas técnicas da funcionalidade Matricula Componente.}
	\label{tabela_arquivo}
	\resizebox{.23\textwidth}{!}{
		\begin{tabular}{|c|c|}
			\hline
			\textbf{A} & \textbf{B} \\ \hline
			\text{GenericSigaaDAO} & \text{0} 	\\ \hline
			\text{ComponenteCurricular} & \text{3} 	\\ \hline
			\text{ComponenteDetalhes} & \text{0} 	\\ \hline
			\text{DiscenteAdapter} & \text{0} 	\\ \hline
			\text{DocenteTurma} & \text{4} 	\\ \hline
			\text{MatriculaComponenteMBean} & \text{0} 	\\ \hline
			\text{SituacaoTurma} & \text{0} 	\\ \hline
			\text{TipoComponenteCurricular} & \text{0} 	\\ \hline
			\text{Turma} & \text{2} 	\\ \hline
			\text{Curriculo} & \text{1} 	\\ \hline
			\text{Discente} & \text{1} 	\\ \hline
		\end{tabular}
	}
\end{table}

Com os valores das dívidas técnicas calculadas para cada funcionalidade é feito a ordenação de forma decrescente, de acordo com o numero de DT presentes nas mesmas.

%!TEX root = ../SBQS_2019_DT.tex

\section{TDVision}
\label{sec:tool}

A partir da definição do método na Seção ~\ref{sec:metodo}, foi desenvolvido um módulo computacional no âmbito deste trabalho, denominado TDVision\footnote{http://easii.ufpi.br/codivision}  e que está incorporado à ferramenta CoDiVision ~\cite{codivision}. Ele tem como objetivo a identificar/visualizar s DT nas funcionalidades de um projeto de \textit{software}, e assim direcionar os desenvolvedores no momento das correções das dívidas.

A TDVision realiza o processo de extração dos dados de repositórios, bem como das métricas para cálculo das Dívidas Técnicas de forma automatizada. Conforme explicado na Seção ~\ref{sec:metodo}, ela utiliza-se da \textit{API} jGit\footnote{https://www.eclipse.org/jgit} para extrair informações dos repositórios de \textit{software}. Para obter os dados sobre as métricas de códigos, \textit{code smells} e DT, ela utiliza-se da \textit{API} RepositoryMiner\footnote{https://github.com/visminer/repositoryminer}.

\renewcommand{\thefigure}{\arabic{figure}}
\begin{figure}[h]
	\begin{center}
		\includegraphics[scale=0.32]{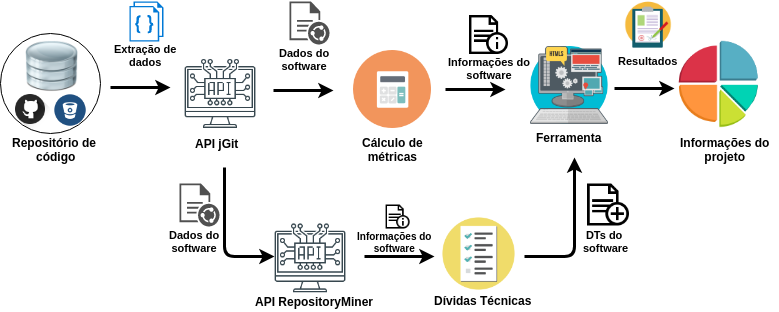}
		\caption{Arquitetura simplificada do TDVision.}
		\label{fig_arquitetura}
	\end{center}
\end{figure}

Na Figura ~\ref{fig_arquitetura} é apresentada a arquitetura simplificada do TDVision. Ela utiliza de serviços existentes da CoDiVision para obter informações do projeto. Após isso, a \textit{API} RepositoryMiner realiza o processo de identificação de \textit{code smells}. Atualmente, são extraídos \textit{code smells} de códigos desenvolvidos na linguagem Java.

\begin{figure}[h]
	\begin{center}
		\includegraphics[scale=0.32]{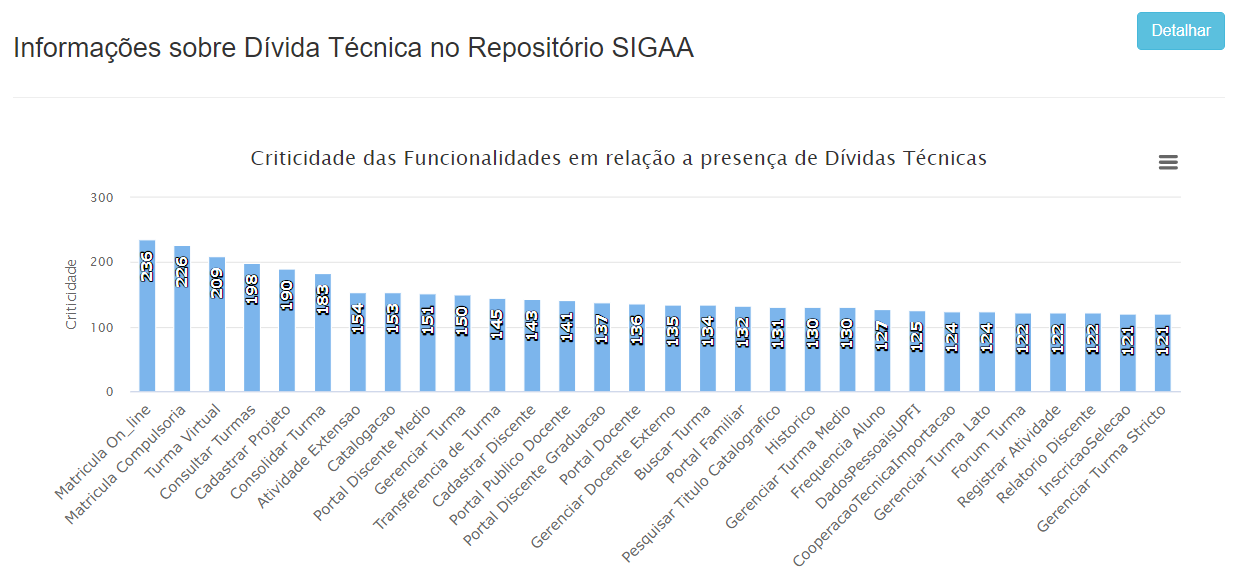}
		\caption{Funcionalidades exibidas na TDVision.}
		\label{fig_tela}
	\end{center}
\end{figure}

Na Figura ~\ref{fig_tela} é apresentado um exemplo de tela da TDVision, na qual, é possível observar as funcionalidades ordenadas de acordo com as quantidades de DT.

\begin{figure}[h]
	\begin{center}
		\includegraphics[scale=0.6]{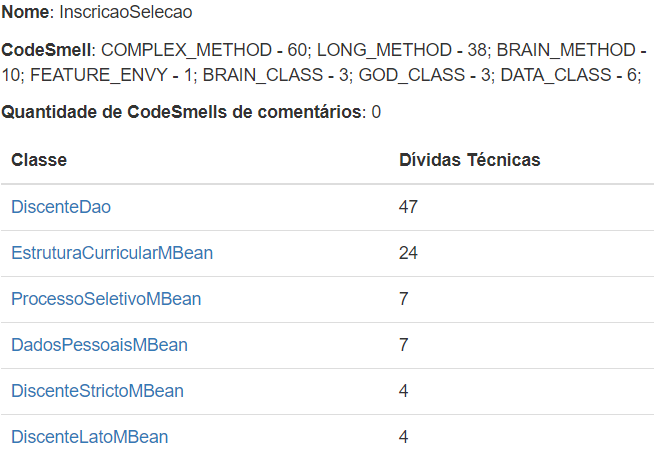}
		\caption{Alguns arquivos que compõem a funcionalidade inscrição Seleção exibidos na TDVision.}
		\label{fig_tela1}
	\end{center}
\end{figure}
Na Figura ~\ref{fig_tela1} é apresentado outro exemplo de tela da TDVision, onde é possível observar informações sobre a funcionalidade, tais como: (i) os tipos de DT presentes na funcionalidade, (ii) a quantidade de cada tipo de DT, (iii) os arquivos que compõem a funcionalidade e suas respectivas quantidades de DT;

\begin{figure}[!h]
\centering
\subfloat[Informações da Classe]{
\includegraphics[scale=0.53]{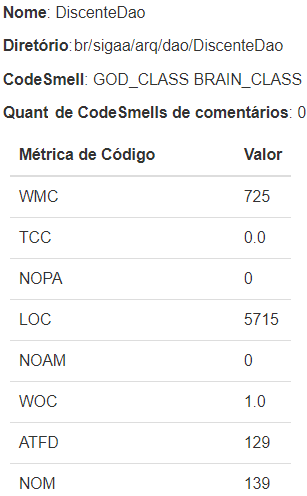}
\label{fig:subfig_a}
}
\subfloat[Informações da Método]{
\includegraphics[scale=0.53]{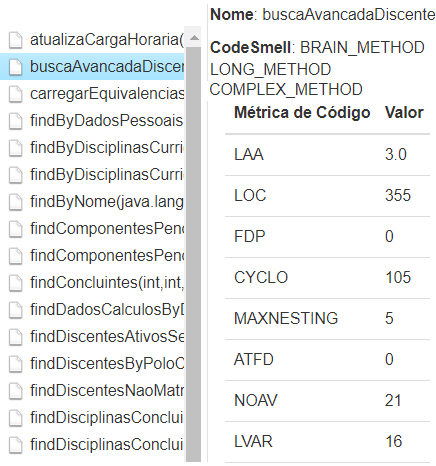}
\label{fig:subfig_b}
}
\caption{Informações de uma classe e de um método que compõem a funcionalidade inscrição Seleção exibidas na TDVision.}
\label{fig:clamet}
\end{figure}

%\gaa{Nada essencial, mas seria melhor se utiliza-se subfigure p/ dividir a figura 7 em duas. O label na própria figura fica com qualidade inferior. Veja o exemplo comentado que deixei no código latex acima.(corrigido)}

Na Figura ~\ref{fig:clamet} é apresentado mais um exemplo de tela da TDVision, onde são exibidos os  \textit{code smells} e os valores obtidos nas métricas de \textit{software}, tanto da classe quanto dos métodos. A Figura ~\ref{fig:subfig_a}, mostra  informações calculadas para uma determinada classe da funcionalidade, e a Figura ~\ref{fig:subfig_b}, exibe informações calculadas de um determinado método de uma classe da funcionalidade.
%!TEX root = ../SBQS_2019_DT.tex

\section{Abordagem Proposta}
\label{sec:abordagem}

A partir da definição do método para identificação de DT por funcionalidade e sua implementação (TDVision), apresentados nas Seções ~\ref{sec:metodo} e ~\ref{sec:tool} respectivamente, foi definida uma abordagem para inserção desse módulo em um ambiente de desenvolvimento real, cujo o principal objetivo é avaliar se a mesma pode estimular os desenvolvedores a corrigirem as DT durante a manutenção do \textit{software}.

A abordagem é projetada para ser flexível, de modo que possa ser adaptada para incorporar novas ideias e abordagens. O usuário abre um chamado, através de uma ferramenta de gerenciamento de projeto/tarefas, que pode ser reportando um erro em uma funcionalidade ou solicitando uma melhoria em uma funcionalidade existente. Esse chamado é então distribuído ao time de desenvolvimento pelo gerente do projeto. Os desenvolvedores de posse das tarefas e seguindo processo de manutenção anterior existente, começariam a trabalhar nas mesmas.

Com a implantação da nossa abordagem o processo de manutenção  ganha novas etapas, ou seja, a partir do momento que os desenvolvedores recebem as tarefas os mesmos utilizam o modulo TDVision, para visualizarem as dívidas técnicas presentes na funcionalidade e quando começam a trabalhar na solução da tarefa também realizam pagamento das DT. A Figura ~\ref{fig:passo_abordagem} mostra os passos da abordagem proposta adotando o modulo TDVision.

\renewcommand{\thefigure}{\arabic{figure}}
\begin{figure}[htpb]
   \centering
   \includegraphics[scale=.6]{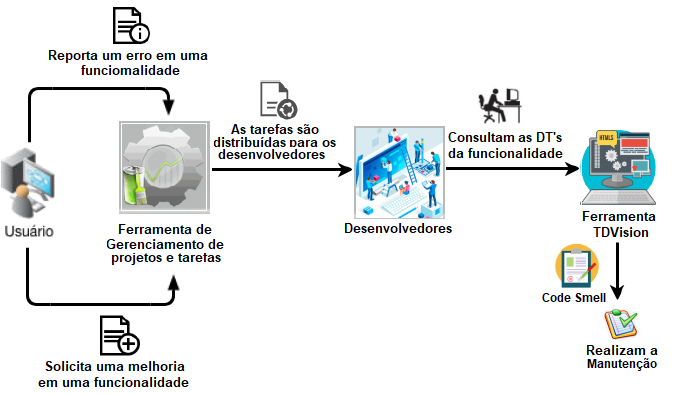}
   \caption{Visão geral dos processos que compõem a abordagem proposta}
   \label{fig:passo_abordagem}
\end{figure}
%!TEX root = ../SBQS_2019_DT.tex
\section{Estudo de Caso}
\label{sec:avaliacao}
\subsection{Introdução}
No campo da engenharia de \textit{software}, a utilização de métodos experimentais pode trazer alguns benefícios, dentre os quais: acelerar o progresso através da rápida eliminação de abordagens não fundamentadas, suposições errôneas e modismos, bem como, ajudar a orientar a engenharia e a teoria para direções promissoras~\cite{Wohlin}. É possível encontrar na literatura três estratégias experimentais: survey, estudo de caso e experimento. Neste sentido, realizamos um estudo de caso  para avaliar a abordagem proposta.

\subsection{Planejamento}

Esta seção apresenta o plano do estudo de caso, destacando seus objetivos, questões, contexto, participantes e procedimentos.

\subsubsection{Objetivo e questão da pesquisa}

O objetivo deste estudo de caso é investigar os benefícios da identificação/visualização de dívidas técnicas em nível de funcionalidade para \textit{softwares} em manutenção/evolução. 

Assim, pretende-se avaliar a seguinte questão geral: 
\begin{itemize}
    \item  A utilização da abordagem que contempla o método de identificação de DT por funcionalidade pode estimular os desenvolvedores a corrigirem as dívidas técnicas durante a manutenção do \textit{software}? 
\end{itemize}

\subsection{Proposições e hipóteses}
A sentença a seguir, baseada no método GQM ~\cite{basili}, define o estudo de caso.

"\textbf{Analisar} a aplicação da abordagem aqui proposta,
\textbf{com o propósito de} avaliar seus benefícios,
\textbf{com respeito ao} gerenciamento de dívida técnica
\textbf{do ponto de vista} de desenvolvedores e gestores de projetos,  
\textbf{no contexto} de projetos de manutenção e evolução de \textit{softwares}”.

Conforme será esclarecido em maiores detalhes na próxima seção, a avaliação da abordagem de acordo com a definição acima será feita a partir da coleta e análise de dados quantitativos e qualitativos durante a operação da avaliação. 

Em um estudo de caso, é necessário estabelecer claramente o que se pretende avaliar ~\cite{Juzgado}. Para isso hipóteses precisam ser definidas. Neste trabalho utilizamos as seguintes hipóteses:

\begin{itemize}

\item $HN_{0}$ :  Não há benefício em relação ao gerenciamento de DT em usar a abordagem proposta durante a manutenção do \textit{software}. 

\[ HN_{0} : \left ( \Delta _{1} = \Delta _{2} \right ) \wedge \left (\Omega _{1} = \Omega _{2}\right )\]

\item $HA_{1}$ : A abordagem proposta produz benefícios no gerenciamento de DT durante a manutenção do \textit{software}. 

\[HA_{1} : \left (\Delta _{2} < \Delta _{1}\right ) \wedge \left (\Omega _{2} > \Omega _{1}\right )\]
\end{itemize}
Onde:

• \textit{$\Delta _{1}$}: Número DT Inseridas pré-abordagem;

• \textit{$\Delta _{2}$}: Número DT Inseridas pós-abordagem;

• \textit{$\Omega _{1}$}: Número DT Corrigidas pré-abordagem;

• \textit{$\Omega _{2}$}: Número DT Corrigidas pós-abordagem;

\subsubsection{Seleção de Variáveis}
O experimento utiliza as seguintes variáveis: 

\begin{itemize}
\item Variável independente: A abordagem utilizada para dar suporte para o pagamento das DT em projetos de \textit{software}.
\item Variáveis dependentes: as quantidades de DT pagas e inseridas durante o uso da abordagem, o processo utilizado, a experiência dos desenvolvedores. 
\end{itemize}

\subsubsection{Conjectura e unidade de análise}

Este estudo foi executado de forma on-line, uma vez que ele aconteceu em tempo real durante um projeto. Os participantes do estudo são profissionais que atuam em um projeto real de manutenção e evolução de \textit{software}. Houve um monitoramento contínuo sobre as atividades dos participantes. 

A unidade de análise definida foi um órgão responsável por Tecnologia da Informação em uma Instituição de Ensino de Superior. O projeto estudado é composto por 764.935 linhas de código (contabilizado somente as escritas em linguagem de programação Java e que não estavam comentadas), 5.335 classes que compõem 982 funcionalidades distribuídas em 18 módulos e 14 portais. Nesse projeto, foram identificadas 4.216 Code Smells (contabilizado somente os 7 tipos que estamos trabalhando nesse estudo: \textit{complex  method} = 1.947, \textit{long method} = 1.436,   \textit{brain method} = 439, \textit{god class} = 150, \textit{brain class} = 139, \textit{data class} = 69 e \textit{feature envy = 36}).

\begin{figure}[h]
	\begin{center}
		\includegraphics[scale=0.4]{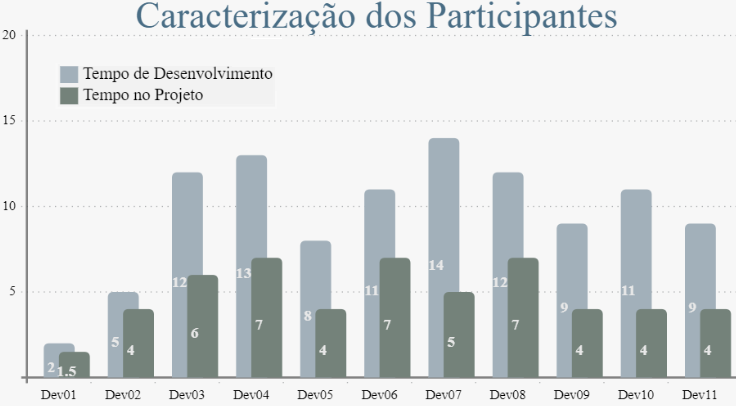}
		\caption{Caracterização dos Participantes}
		\label{fig_caracterizacao}
	\end{center}
\end{figure}

Os sujeitos deste experimento foram selecionados por amostragem por conveniência ~\cite{Wohlin}, ou seja, não foram escolhidos de forma aleatória. Este fato otimizou o tempo e o custo da realização do estudo.  Dessa forma, os participantes selecionados são o gerente do projeto e os analistas de desenvolvimento que trabalham no projeto onde será executado o estudo.
O perfil dos participantes em termos de sua experiência prática com desenvolvimento de \textit{software} e tempo no projeto foi levantando através da aplicação de um questionário. A Figura ~\ref{fig_caracterizacao} indica a caracterização dos participantes em termos de experiência em desenvolvimento de \textit{software} e tempo de participação no projeto em que realizou-se o estudo.

\subsubsection{Métodos de coleta dados}

Foram utilizados diferentes fontes de dados para a realização da coleta de dados antes e depois da execução do estudo de caso, listados a seguir:

\begin{itemize}
\item \textbf{Registros de realização de tarefas:} a contabilização da quantidade de tarefas
realizadas pela equipe de desenvolvimento, assim como o tempo gasto na execução dessas tarefas foram obtidas a partir da ferramenta de registro de tarefa existente no orgão, tendo sido obtidos dados referentes a períodos anteriores à realização do estudo de caso e os dados obtidos durante a aplicação da abordagem;
 
\item \textbf{Repositórios de código de fonte dos \textit{software} desenvolvidos:} a contabilização das DT inseridas e pagas foram obtidas minerando o repositório de código de fonte do orgão.
Esses dados também estavam relacionados a períodos anteriores ao estudo de caso,
bem como durante a realização do estudo.

\item \textbf{Questionários aplicados aos participantes:} um questionário de caracterização dos participantes e outro de \textit{feedback}, para possibilitar a análise qualitativa;
\end{itemize}

\subsection{Operação}

Inicialmente, os participantes receberam um treinamento com duração de duas horas e trinta minutos, em que foi explicado a finalidade do estudo, os objetivos da abordagem, a descrição do cenário da atividade proposta. Em seguida foi solicitado aos participantes que preenchessem os formulários de consentimento e caracterização. Então, uma primeira rodada foi utilizada para que os participantes se familiarizassem com a tarefa de refatoração a ser realizada, servindo apenas como uma \textit{baseline}. O estudo em si, foi realizado durante o período de 04/12/2018 a 25/04/2019, em que alguns dados foram coletados, tais como duração de atividades relacionadas ao gerenciamento da dívida técnica, total de DT inseridas e pagas.  

Para o cadastro e o acompanhamento das tarefas, sejam elas evolutivas (customizações) ou corretivas (sustentações), o órgão utiliza um \textit{software} próprio que foi desenvolvido pelo time, denominado de SINAPSE (Sistema Integrado de Acompanhamento a Projetos e Serviços). Esse sistema é integrado ao Redmine, que é uma aplicação web flexível utilizada para gerenciamento de projetos. O processo de trabalho da equipe para customizações é elaborado de acordo com o \textit{framework Scrum}. As \textit{sprints} no órgão tem a duração de 10 dias úteis e possuem alguns passos bem definidos. 

A ferramenta descrita na Seção ~\ref{sec:tool} foi inserida no ambiente de desenvolvimento do orgão e operou neste ambiente durante vinte semanas. Durante esse período  as tarefas que estavam relacionadas a melhorias em funcionalidades já existentes e as que reportavam erros em uma funcionalidade, tinham suas DT checadas na TDVision, pelo desenvolvedor, antes de começar a trabalhar na tarefa. Dessa forma os desenvolvedores obtinham conhecimento dos tipos de dívidas encontradas na funcionalidade e em quais arquivos da funcionalidade elas estavam presentes, podendo assim realizar o pagamento das mesmas.   

É importante ressaltar a dificuldade e o cuidado necessário nesse estudo, por conta de ser realizado no contexto de um projeto de desenvolvimento de \textit{software} real e com a equipe efetiva deste projeto. Uma dificuldade era o fator tempo/produtividade, pois o time de desenvolvimento, baseado em retrospectivas anteriores, tinha uma capacidade de entregar até 20 pontos de função por \textit{sprint}, e com as tarefas extras de refatoração para pagamento de DT, havia a preocupação de que ocorresse a diminuição da produtividade. Esta questão foi tratada limitando o tempo para realizar a tarefa de refatoração para pagamento das DT em 2 horas, sendo executada antes do início da tarefa de manutenção do \textit{software}. 

A cada 2 semanas, eram realizadas extrações no repositório de código, coletando o número de DT inseridas e pagas, e na ferramenta de acompanhamento das tarefas eram coletadas informações sobre o total de tarefas e o tempo de execução das mesmas. No final do estudo, os participantes foram solicitados a preencher um questionário de acompanhamento, permitindo obter dados qualitativos.

\subsection{Resultados e discussão}

Após a coleta de dados, na fase de operação, estes devem ser analisados e interpretados para que se possa tirar conclusões~\cite{Wohlin}.

\subsubsection{Análise quantitativa}
A análise deste estudo tem por objetivo comparar os dados coletados no período de execução do experimento com o período anterior ao mesmo.

\begin{figure}[h]
	\begin{center}
		\includegraphics[scale=0.4]{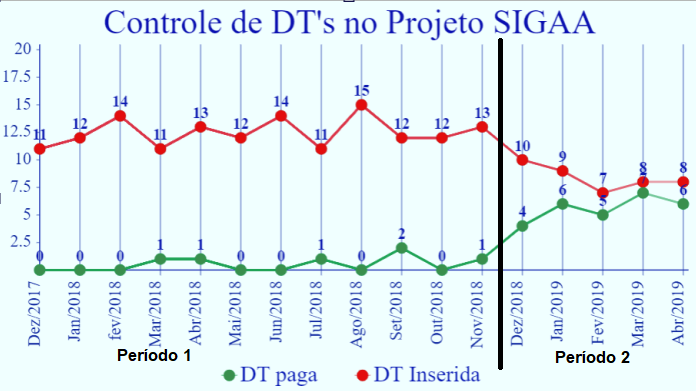}
		\caption{Controle das DT no projeto}
		\label{fig_controle}
	\end{center}
\end{figure}

A Figura ~\ref{fig_controle} mostra uma comparação do período 1 (antes da aplicação da abordagem), com o período 2 em que foi realizado o estudo, mostrando o número de novas DT inseridas e o número de DT pagas.

Como é possível observar na comparação do período 1 com o período 2, com a utilização do uso da abordagem proposta houve uma melhora no tratamento das DT: (i) aumento do número de DT pagas e (ii) diminuição do número de DTs inseridas.

\begin{figure}[h]
	\begin{center}
		\includegraphics[scale=0.4]{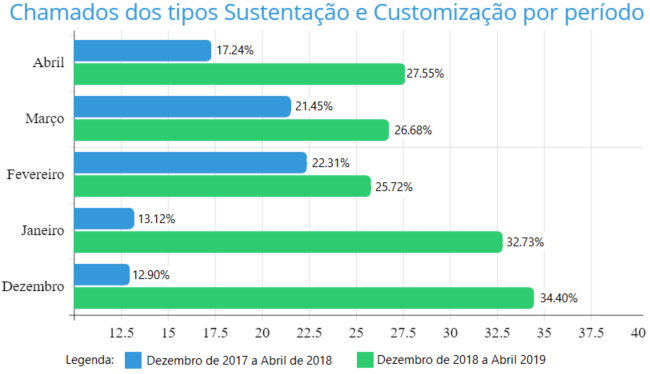}
		\caption{Chamados dos tipos sustentação e customização por período}
		\label{fig_chamado_total}
	\end{center}
\end{figure}

\begin{figure}[h]
	\begin{center}
		\includegraphics[scale=0.45]{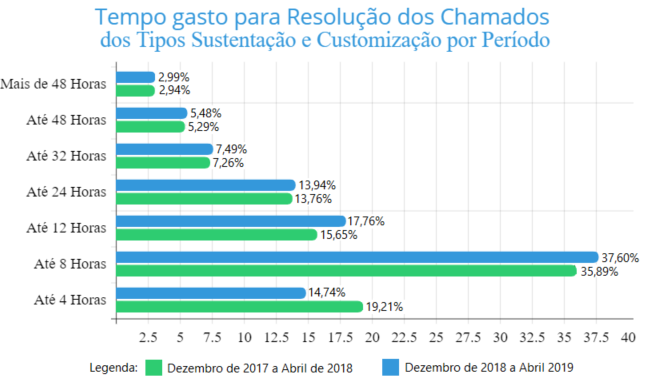}
		\caption{Tempo gasto para resolução dos chamados}
		\label{fig_chamados_hora}
	\end{center}
\end{figure}

Adicionalmente, realizamos um levantamento do total de chamados  abertos, e do tempo gasto pelo time para resolvê-los para com isso podermos comparar com o período em que o estudo foi realizado. Na Figura ~\ref{fig_chamado_total}, é possível observar que o número de chamados é maior no período de dezembro de 2018 a abril de 2019, e mesmo assim o numero de DT inseridas foi menor quando comparado ao período de dezembro de 2017 a abril de 2018, conforme Figura ~\ref{fig_controle}. Com relação à Figura  ~\ref{fig_chamados_hora}, podemos observar que houve uma redução na resolução dos chamados em até 4 horas, e consequentemente houve um aumento significativo nos chamados resolvidos em até 8 horas e em até 12 horas, o que é perfeitamente normal tendo em vista que os desenvolvedores passaram gastar tempo refatorando para pagarem as DT.

Usando o \textit{CausalImpact}\footnote{https://github.com/google/CausalImpact}\footnote{https://google.github.io/CausalImpact/CausalImpact.html}, que é um algorítimo desenvolvido pelo Google para estimar o efeito causal de uma intervenção projetada em uma série temporal, realizamos uma projeção, baseada no histórico, de como seria caso o ambiente de desenvolvimento não tivesse sido alterado e comparamos estatística essa projeção com os dados obtidos no estudo. O \textit{CausalImpact} funciona da seguinte forma: dada uma série temporal de resposta (no nosso caso a inserção e correção das DT pós implantação da abordagem) e um conjunto de séries temporais de controle (inserção e correção das DT pré implantação da abordagem), o pacote constrói um modelo bayesiano de séries temporais estruturais. Este modelo é então usado para tentar prever o contrafactual, ou seja, como a métrica de resposta teria evoluído após a intervenção se a intervenção nunca tivesse ocorrido.

\begin{figure}[h]
	\begin{center}
		\includegraphics[scale=0.65]{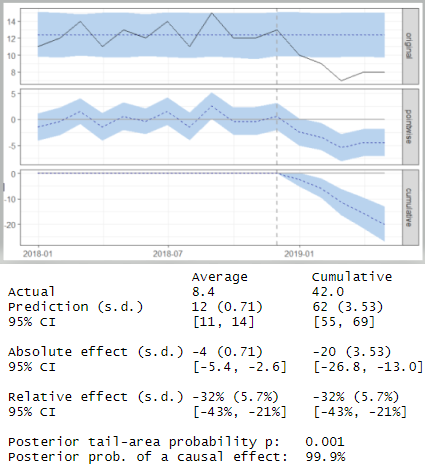}
		\caption{Gráfico das DT inseridas}
		\label{fig_inseridas}
	\end{center}
\end{figure}

\begin{figure}[h]
	\begin{center}
		\includegraphics[scale=0.65]{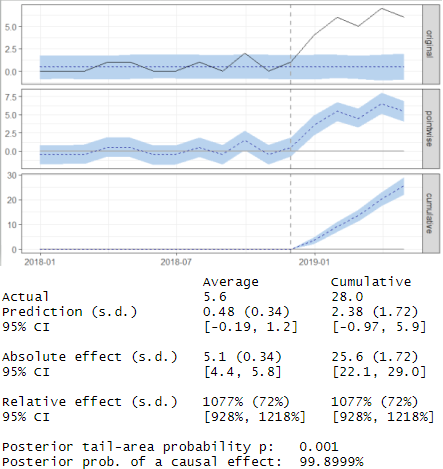}
		\caption{Gráfico das DT Corigidas}
		\label{fig_corigidas}
	\end{center}
\end{figure}

Por padrão, o gráfico gerado pelo \textit{CausalImpact} contém três painéis, conforme é possível observar nas Figuras ~\ref{fig_inseridas} e ~\ref{fig_corigidas}. O primeiro painel mostra os dados e uma previsão contrafactual para o período pós-aplicação da abordagem. O segundo painel mostra a diferença entre dados observados e previsões contrafactuais. Este é o efeito causal pontual, conforme estimado pelo modelo. O terceiro painel adiciona as contribuições pontuais do segundo painel, resultando em um gráfico do efeito cumulativo da aplicação da abordagem.

Conforme Figura ~\ref{fig_inseridas}, durante o período após aplicação da abordagem,  a variável resposta apresentou um valor médio de aproximadamente 8,40. Em contraste, na ausência de uma intervenção, era esperado uma resposta média de 12,41. O intervalo de 95\% desta previsão contrafactual está entre 11,01 e 13,77. A variável de resposta teve um valor global de 42,00. Em contraste, se a aplicação da abordagem não tivesse ocorrido, teríamos esperado uma soma de 62,06. Ou seja, era esperado um valor dentro do intervalo de 55,04 e 68.84, caso a abordagem não tivesse sido aplicada.

Os resultados anteriores são dados em termos de números absolutos. Em termos relativos, a variável resposta mostrou uma diminuição de 32\%. A probabilidade de obter este efeito por acaso é muito pequena (probabilidade bayesiana de área de cauda unilateral p = 0,001). Isso significa que o efeito causal, provocado pela aplicação da abordagem, pode ser considerado estatisticamente significativo.

Analisando a Figura ~\ref{fig_corigidas}, é possível observar que durante o período pós-aplicação da abordagem, a variável resposta, que representa o número de DT pagas, apresentou um valor médio de aproximadamente 5,60. Em contraste, na ausência de uma intervenção, esperávamos uma resposta média de 0,48. Ou seja, era esperado um valor dentro do intervalo de -0,19 e 1,18. Subtrair essa previsão da resposta observada produz uma estimativa do efeito causal da aplicação da abordagem sobre a variável resposta. Este efeito é de 5,12. A variável resposta teve um valor global de 28,00. Em contraste, se a aplicação da abordagem não tivesse ocorrido, teríamos esperado uma soma de 2,38. 

Os resultados acima são dados em termos de números absolutos. Em termos relativos, a variável resposta apresentou um aumento de  1077\%. Isto significa que o efeito positivo observado durante o período de aplicação da abordagem é estatisticamente significativo e é improvável que seja devido a flutuações aleatórias. A probabilidade de obter este efeito por acaso é muito pequena (probabilidade bayesiana de área de cauda unilateral p = 0,001).

Esses resultados nos permite refutar a hipótese nula $HN_{0}$ e, ao mesmo tempo, validar a hipótese alternativa $HA_{1}$. 

\subsubsection{Análise qualitativa}

Os dados dos sujeitos que usaram a abordagem foram analisados qualitativamente. A opinião foi coletada a partir do questionário de feedback, aplicado logo após a execução do experimento. A escala utilizada para cada uma das questões é do tipo \textit{Likert}~\cite{likert}. No nosso estudo empregou-se os seguintes níveis de afirmação na escala de Likert: Discordo plenamente; Discordo parcialmente; Nem concordo, nem discordo; Concordo parcialmente; Concordo plenamente. As seguintes questões foram feitas: (i) O uso da abordagem ajudou a melhorar a qualidade do seu código produzido?, (ii) O esforço para realização das tarefas que tinham nota informando a presença de dívidas técnicas foi diferente das que não tinham?, (iii) O fato de ter conhecimento sobre a presença de DT afetou de alguma forma como você desenvolve \textit{software}?, (iv) Você acha importante monitorar dívida técnica dentro do processo de desenvolvimento de \textit{software}? e (v) É possível fazer com que este gerenciamento faça parte do processo de desenvolvimento? 

 As respostas foram analisadas e sumarizadas a seguir: 81.82\% dos participantes notaram um aumento na qualidade do seu código produzido.  No entanto, 90.91\% notaram um aumento no esforço para realizar a tarefa, pois além de terem que corrigir um defeito, sem inserir novas DT, eles tinham que refatorar para pagar as DT já existentes. Eles confirmaram de forma unânime  que o conhecimento sobre a presença de DT afetou o modo como eles desenvolviam. E 81.82\% acham importante monitorar dívida técnica dentro do processo de desenvolvimento de \textit{software} e quanto a abordagem fazer parte do processo de desenvolvimento, 72.73\% dos entrevistados consideraram que é possível.

%!TEX root = ../SBQS_2019_DT.tex

\section{Ameaças à Validade} 
\label{sec:ameacas}

Considerações a respeito da validade interna e externa  do estudo se encontram descritas a seguir:  

\textbf{Validade Interna}: Uma ameaça interna identificada é a interação entre os participantes no momento de responder o questionário, uma vez que o mesmo foi submetido aos participantes em um ambiente não controlado e em momentos distintos, por isso, não é possível afirmar que os participantes não interagiram entre si com o intuito de compartilhar informações sobre as suas respostas, oferecendo uma ameaça à autenticidade das respostas obtidas. Outra ameaça identificada é a classificação de experiência dos participantes pois, o fato de um participante está com certo tempo trabalhando com uma linguagem ou em um projeto não garante que eles tenham um grande conhecimento sobre os mesmos, além disso nas respostas do questionário eles podem relatar o que desejam ser e não o que de fato são. Isso poderia ser substituído pela execução de avaliações de habilidades dos participantes com custo de esforço e tempo maior para a realização do estudo. No entanto, o mesmo número de aspectos do histórico dos participantes poderiam permanecer incerto e ocorrer uma evasão no número de participantes, tendo em vista que os mesmos poderiam ter receio de represália por expor seu conhecimento quanto ao seu trabalho realizado.

\textbf{Validade Externa} : Apesar de o estudo ter sido feito em um sistema de grande porte, por ser baseado em apenas um projeto, os resultados podem não refletir outros ambientes de desenvolvimento que utilizem outras tecnologias. Também estamos cientes que limitamos a nossa atenção apenas aos códigos escritos em linguagem programação Java, devido às limitações da infra-estrutura que usamos (por exemplo,  a abordagem de identificação de funcionalidade e a ferramenta de detecção de \textit{code smell} que só funcionam em código Java). São desejáveis mais estudos que visem replicar o nosso trabalho sobre sistemas escritos em outras linguagens de programação.

%!TEX root = ../SBQS_2019_DT.tex

\section{Conclusão}
\label{sec:conclusao}

Neste trabalho foi apresentada um método, para identificação/visualização de DT sobre uma nova perspectiva: funcionalidade de \textit{software}.  O método faz uso das técnicas de Mineração de Repositório de \textit{software} para identificar as funcionalidades e em seguida as DT presentes nas mesmas. 

Além do método foi apresentado um novo módulo da ferramenta CoDiVision, o TDVision que foi construído a partir do método proposto nesse trabalho. Com ele, as atividades de identificação e monitoramento de dívidas técnicas podem ser realizadas de forma simples e automatizada. Também apresentamos uma abordagem para manutenção/evolução de \textit{software} que faz uso do modulo TDVision e tem como foco a redução de DT;

Um estudo de caso foi realizado em um projeto real para verificar os benefícios da identificação/visualização do DT no nível de funcionalidade. Os resultados combinados com os relatos dos entrevistados perante as questões, mostram evidências de que a abordagem pode apoiar a gestão técnica da dívida podendo-se assim oferecer amparo na tomada de decisão quanto à negociação com \textit{stakeholders}, auxiliando a comunicar melhor a dívida aos envolvidos no processo de desenvolvimento de \textit{software}.
 
Como proposta futura, temos como objetivo  estender a detecção considerando outros tipos de DT e como nossa abordagem é projetada para ser flexível, de modo que possa ser adaptada para incorporar novas ideias e abordagens, é importante realizar um novo estudo aplicando o método de identificação por funcionalidade, proposto nesse trabalho, e um outro método que identifica DT por arquivo, dividindo a equipe(onde uma parte usa o método que identifica por funcionalidade e outra parte usa uma outra ferramenta que identifica por arquivo) e assim comparar qual método tem melhores resultados (tanto no gerenciamento de DT quanto na adaptação ao ambiente de desenvolvimento e facilidade de uso);

\bibliographystyle{ACM-Reference-Format}
\bibliography{bibliografia}


\begin{thebibliography}{34}

%%% the Web address.

\ifx \showCODEN    \undefined \def \showCODEN     #1{\unskip}     \fi
\ifx \showDOI      \undefined \def \showDOI       #1{#1}\fi
\ifx \showISBNx    \undefined \def \showISBNx     #1{\unskip}     \fi
\ifx \showISBNxiii \undefined \def \showISBNxiii  #1{\unskip}     \fi
\ifx \showISSN     \undefined \def \showISSN      #1{\unskip}     \fi
\ifx \showLCCN     \undefined \def \showLCCN      #1{\unskip}     \fi
\ifx \shownote     \undefined \def \shownote      #1{#1}          \fi
\ifx \showarticletitle \undefined \def \showarticletitle #1{#1}   \fi
\ifx \showURL      \undefined \def \showURL       {\relax}        \fi
% invisible to TeX
\providecommand\bibfield[2]{#2}
\providecommand\bibinfo[2]{#2}
\providecommand\natexlab[1]{#1}
\providecommand\showeprint[2][]{arXiv:#2}

\bibitem[\protect\citeauthoryear{Alves, Mendes, de~Mendon{\c{c}}a,
  Sp{\'\i}nola, Shull, and Seaman}{Alves et~al\mbox{.}}{2016}]%
        {alves2016identification}
\bibfield{author}{\bibinfo{person}{Nicolli~SR Alves}, \bibinfo{person}{Thiago~S
  Mendes}, \bibinfo{person}{Manoel~G de Mendon{\c{c}}a},
  \bibinfo{person}{Rodrigo~O Sp{\'\i}nola}, \bibinfo{person}{Forrest Shull},
  {and} \bibinfo{person}{Carolyn Seaman}.} \bibinfo{year}{2016}\natexlab{}.
\newblock \showarticletitle{Identification and management of technical debt: A
  systematic mapping study}.
\newblock \bibinfo{journal}{\emph{Information and Software Technology}}
  \bibinfo{volume}{70} (\bibinfo{year}{2016}), \bibinfo{pages}{100--121}.
\newblock


\bibitem[\protect\citeauthoryear{Basili}{Basili}{1999}]%
        {basili}
\bibfield{author}{\bibinfo{person}{Forrest Shull Filippo~Lanubile. Basili,
  Victor~R.}} \bibinfo{year}{1999}\natexlab{}.
\newblock \showarticletitle{Building Knowledge through Families of
  Experiments.}
\newblock \bibinfo{journal}{\emph{IEEE Transactions on Software Engineering}}
  (\bibinfo{year}{1999}).
\newblock


\bibitem[\protect\citeauthoryear{Campbell and Papapetrou}{Campbell and
  Papapetrou}{2013}]%
        {campbell2013sonarqube}
\bibfield{author}{\bibinfo{person}{G Campbell} {and}
  \bibinfo{person}{Patroklos~P Papapetrou}.} \bibinfo{year}{2013}\natexlab{}.
\newblock \bibinfo{booktitle}{\emph{SonarQube in action}}.
\newblock \bibinfo{publisher}{Manning Publications Co.}
\newblock


\bibitem[\protect\citeauthoryear{Cunningham}{Cunningham}{1992}]%
        {Cunningham:1992:WPM:157709.157715}
\bibfield{author}{\bibinfo{person}{Ward Cunningham}.}
  \bibinfo{year}{1992}\natexlab{}.
\newblock \showarticletitle{The WyCash Portfolio Management System}. In
  \bibinfo{booktitle}{\emph{Addendum to the Proceedings on Object-oriented
  Programming Systems, Languages, and Applications (Addendum)}}
  \emph{(\bibinfo{series}{OOPSLA '92})}. \bibinfo{publisher}{ACM},
  \bibinfo{address}{New York, NY, USA}, \bibinfo{pages}{29--30}.
\newblock
\showISBNx{0-89791-610-7}
\urldef\tempurl%
\url{https://doi.org/10.1145/157709.157715}
\showDOI{\tempurl}


\bibitem[\protect\citeauthoryear{E.~Tom and Vidgen}{E.~Tom and Vidgen}{2013}]%
        {tom}
\bibfield{author}{\bibinfo{person}{A.~Aurum E.~Tom} {and} \bibinfo{person}{R.
  Vidgen}.} \bibinfo{year}{2013}\natexlab{}.
\newblock \showarticletitle{An exploration of technical debt}.
\newblock \bibinfo{journal}{\emph{Journal of Systems and Software}}
  (\bibinfo{year}{2013}).
\newblock


\bibitem[\protect\citeauthoryear{et~al.}{et~al.}{2010}]%
        {brown}
\bibfield{author}{\bibinfo{person}{N.~Brown et al.}}
  \bibinfo{year}{2010}\natexlab{}.
\newblock \showarticletitle{Managing technical debt in software-reliant
  systems}.
\newblock \bibinfo{journal}{\emph{Proceedings of the FSE/SDP workshop on Future
  of software engineering research}} (\bibinfo{year}{2010}).
\newblock


\bibitem[\protect\citeauthoryear{F.~Arcelli and Maggioni.}{F.~Arcelli and
  Maggioni.}{2008}]%
        {marple}
\bibfield{author}{\bibinfo{person}{M.~Zanoni F.~Arcelli, C.~Tosi} {and}
  \bibinfo{person}{S. Maggioni.}} \bibinfo{year}{2008}\natexlab{}.
\newblock \showarticletitle{The marple project: A tool for design pattern
  detection and software architecture reconstruction.}
\newblock \bibinfo{journal}{\emph{In 1st International Workshop on Academic
  Software Development Tools and Techniques (WASDeTT-1)}}
  (\bibinfo{year}{2008}).
\newblock


\bibitem[\protect\citeauthoryear{F.~Vanderson M.~A. and Ibiapina}{F.~Vanderson
  M.~A. and Ibiapina}{2018}]%
        {vanderson:2018}
\bibfield{author}{\bibinfo{person}{Werney A. L.~Lira F.~Vanderson M.~A., Pedro
  A. S.~Neto} {and} \bibinfo{person}{Irvayne M.~S. Ibiapina}.}
  \bibinfo{year}{2018}\natexlab{}.
\newblock \showarticletitle{Analysis of Code Familiarity in Module and
  Functionality Perspectives}.
\newblock \bibinfo{journal}{\emph{SBQS 2018: XVII Simpósio Brasileiro de
  Qualidade de Software}} (\bibinfo{year}{2018}).
\newblock


\bibitem[\protect\citeauthoryear{G, Novais, Gon{\c{c}}alves, Sp{\i}nola,
  Mendon{\c{c}}a, and Salvador}{G et~al\mbox{.}}{2015}]%
        {repositoryMiner}
\bibfield{author}{\bibinfo{person}{Felipe G}, \bibinfo{person}{Thiago~S Novais,
  Mendes}, \bibinfo{person}{Renato Gon{\c{c}}alves, Renato~Novais},
  \bibinfo{person}{Rodrigo~O Sp{\i}nola}, \bibinfo{person}{Manoel
  Mendon{\c{c}}a}, {and} \bibinfo{person}{BA Salvador}.}
  \bibinfo{year}{2015}\natexlab{}.
\newblock \showarticletitle{RepositoryMiner- : uma ferramenta extensível de
  mineração de repositórios de software para identificação automática de
  Dívidas Técnicas}.
\newblock  (\bibinfo{year}{2015}).
\newblock


\bibitem[\protect\citeauthoryear{H.~Humanes and Díaz}{H.~Humanes and
  Díaz}{2017}]%
        {tedma}
\bibfield{author}{\bibinfo{person}{C.~Fernández-Sánchez H.~Humanes,
  J.~Garbajosa} {and} \bibinfo{person}{J. Díaz}.}
  \bibinfo{year}{2017}\natexlab{}.
\newblock \showarticletitle{An open tool for assisting in technical debt
  management. In Software Engineering and Advanced Applications (SEAA)}.
\newblock \bibinfo{journal}{\emph{IEEE, 43rd Euromicro Conference on, pages
  400–403}} (\bibinfo{year}{2017}).
\newblock


\bibitem[\protect\citeauthoryear{Hassan}{Hassan}{2008}]%
        {hassan}
\bibfield{author}{\bibinfo{person}{A.~E. Hassan}.}
  \bibinfo{year}{2008}\natexlab{}.
\newblock \showarticletitle{The road ahead for mining software repositories}.
\newblock \bibinfo{journal}{\emph{In Frontiers of Software Maintenance}}
  (\bibinfo{year}{2008}), \bibinfo{pages}{48--57}.
\newblock


\bibitem[\protect\citeauthoryear{Hemmati, Nadi, Baysal, Kononenko, Wang,
  Holmes, and Godfrey}{Hemmati et~al\mbox{.}}{2013}]%
        {hemmati2013msr}
\bibfield{author}{\bibinfo{person}{Hadi Hemmati}, \bibinfo{person}{Sarah Nadi},
  \bibinfo{person}{Olga Baysal}, \bibinfo{person}{Oleksii Kononenko},
  \bibinfo{person}{Wei Wang}, \bibinfo{person}{Reid Holmes}, {and}
  \bibinfo{person}{Michael~W Godfrey}.} \bibinfo{year}{2013}\natexlab{}.
\newblock \showarticletitle{The msr cookbook: Mining a decade of research}. In
  \bibinfo{booktitle}{\emph{Mining Software Repositories (MSR), 2013 10th IEEE
  Working Conference on}}. IEEE, \bibinfo{pages}{343--352}.
\newblock


\bibitem[\protect\citeauthoryear{Holvitie and Leppänen.}{Holvitie and
  Leppänen.}{2013}]%
        {Debtflag}
\bibfield{author}{\bibinfo{person}{J. Holvitie} {and} \bibinfo{person}{V.
  Leppänen.}} \bibinfo{year}{2013}\natexlab{}.
\newblock \showarticletitle{Debtflag: Technical debt management with a
  development environment integrated tool.}
\newblock \bibinfo{journal}{\emph{In Managing Technical Debt (MTD), 2013 4th
  International Workshop on, pages 20–27. IEEE, 2013.}}
  (\bibinfo{year}{2013}).
\newblock


\bibitem[\protect\citeauthoryear{Juzgado and Vegas}{Juzgado and Vegas}{2004}]%
        {Juzgado}
\bibfield{author}{\bibinfo{person}{Moreno A.~M. Juzgado, N.~J.} {and}
  \bibinfo{person}{S. Vegas}.} \bibinfo{year}{2004}\natexlab{}.
\newblock \showarticletitle{Reviewing 25 Years of Testing Technique
  Experiments.}
\newblock \bibinfo{journal}{\emph{Empirical Software Engineering,9(1-2)}}
  (\bibinfo{year}{2004}), \bibinfo{pages}{7--44}.
\newblock


\bibitem[\protect\citeauthoryear{K.}{K.}{1999}]%
        {FOWLER}
\bibfield{author}{\bibinfo{person}{M.~FOWLER K., BECK}.}
  \bibinfo{year}{1999}\natexlab{}.
\newblock \showarticletitle{Refactoring: improving the design of existing
  code}.
\newblock \bibinfo{journal}{\emph{Addison-Wesley Professional}}
  (\bibinfo{year}{1999}).
\newblock


\bibitem[\protect\citeauthoryear{KAGDI}{KAGDI}{2007}]%
        {KAGDI:2007}
\bibfield{author}{\bibinfo{person}{J.~I.; SHARIF-B. KAGDI, H.;~MALETIC}.}
  \bibinfo{year}{2007}\natexlab{}.
\newblock \showarticletitle{Mining software repositories for traceability
  links.}
\newblock \bibinfo{journal}{\emph{15th IEEE International Conference on Program
  Comprehension,ICPC ’07. doi: 10.1109/ICPC.2007.28}} (\bibinfo{year}{2007}),
  \bibinfo{pages}{145–--154}.
\newblock


\bibitem[\protect\citeauthoryear{KERIEVSKY}{KERIEVSKY}{2005}]%
        {KERIEVSKY}
\bibfield{author}{\bibinfo{person}{KERIEVSKY}.}
  \bibinfo{year}{2005}\natexlab{}.
\newblock \showarticletitle{Refactoring to patterns}.
\newblock \bibinfo{journal}{\emph{Pearson Deutschland GmbH}}
  (\bibinfo{year}{2005}).
\newblock


\bibitem[\protect\citeauthoryear{Kersten and Murphy}{Kersten and
  Murphy}{2005}]%
        {Kersten2005}
\bibfield{author}{\bibinfo{person}{M Kersten} {and} \bibinfo{person}{Gail~C
  Murphy}.} \bibinfo{year}{2005}\natexlab{}.
\newblock \showarticletitle{{Mylar: a degree-of-interest model for IDEs}}.
\newblock \bibinfo{journal}{\emph{4th international conference on
  Aspectoriented software development (AOSD)}} (\bibinfo{year}{2005}),
  \bibinfo{pages}{159--168}.
\newblock


\bibitem[\protect\citeauthoryear{Lanza and Marinescu}{Lanza and
  Marinescu}{2007}]%
        {lanza2007object}
\bibfield{author}{\bibinfo{person}{Michele Lanza} {and} \bibinfo{person}{Radu
  Marinescu}.} \bibinfo{year}{2007}\natexlab{}.
\newblock \bibinfo{booktitle}{\emph{Object-oriented metrics in practice: using
  software metrics to characterize, evaluate, and improve the design of
  object-oriented systems}}.
\newblock \bibinfo{publisher}{Springer Science \& Business Media}.
\newblock


\bibitem[\protect\citeauthoryear{Martini and Bosch}{Martini and Bosch}{2017}]%
        {anacodedebt}
\bibfield{author}{\bibinfo{person}{A. Martini} {and} \bibinfo{person}{J.
  Bosch}.} \bibinfo{year}{2017}\natexlab{}.
\newblock \showarticletitle{The magnificent seven: towards a systematic
  estimation of technical debt interest.}
\newblock \bibinfo{journal}{\emph{In Proceedings of the XP2017 Scientific
  Workshops, page 7. ACM,}} (\bibinfo{year}{2017}).
\newblock


\bibitem[\protect\citeauthoryear{McConnell}{McConnell}{2007}]%
        {mcconnell2007technical}
\bibfield{author}{\bibinfo{person}{Steve McConnell}.}
  \bibinfo{year}{2007}\natexlab{}.
\newblock \showarticletitle{Technical debt}.
\newblock \bibinfo{journal}{\emph{Software Best Practices, Nov}}
  (\bibinfo{year}{2007}).
\newblock


\bibitem[\protect\citeauthoryear{McIver}{McIver}{1991}]%
        {likert}
\bibfield{author}{\bibinfo{person}{E.~G. McIver, J. P.and~Carmines}.}
  \bibinfo{year}{1991}\natexlab{}.
\newblock \showarticletitle{Unidimensional Scaling.}
\newblock \bibinfo{journal}{\emph{Sage Publications.}} (\bibinfo{year}{1991}).
\newblock


\bibitem[\protect\citeauthoryear{Mendes, Novais, Mendonca, Carvalho, and
  Gomes}{Mendes et~al\mbox{.}}{2017}]%
        {170925}
\bibfield{author}{\bibinfo{person}{Thiago Mendes}, \bibinfo{person}{Renato
  Novais}, \bibinfo{person}{Manoel Mendonca}, \bibinfo{person}{Luis Carvalho},
  {and} \bibinfo{person}{Felipe Gomes}.} \bibinfo{year}{2017}\natexlab{}.
\newblock \showarticletitle{RepositoryMiner - uma ferramenta extensível de
  mineração de repositórios de software para identificação automática de
  Dívida Técnica}. In \bibinfo{booktitle}{\emph{CBSoft 2017 - Sessao de
  Ferramentas ()}}.
\newblock
\urldef\tempurl%
\url{http://XXXXX/170925.pdf}
\showURL{%
\tempurl}


\bibitem[\protect\citeauthoryear{Mendes, Gon{\c{c}}alves, Gomes, Novais,
  Sp{\i}nola, Mendon{\c{c}}a, and Salvador}{Mendes et~al\mbox{.}}{2015}]%
        {mendesvisminertd}
\bibfield{author}{\bibinfo{person}{Thiago~S Mendes}, \bibinfo{person}{David~P
  Gon{\c{c}}alves}, \bibinfo{person}{Felipe~G Gomes}, \bibinfo{person}{Renato
  Novais}, \bibinfo{person}{Rodrigo~O Sp{\i}nola}, \bibinfo{person}{Manoel
  Mendon{\c{c}}a}, {and} \bibinfo{person}{BA Salvador}.}
  \bibinfo{year}{2015}\natexlab{}.
\newblock \showarticletitle{VisminerTD: Uma Ferramenta para Identifica{\c{c}}
  ao Autom{\'a}tica e Monitoramento Interativo de D{\i}vida T{\'e}cnica}.
\newblock  (\bibinfo{year}{2015}).
\newblock


\bibitem[\protect\citeauthoryear{Moura, Lira, Ibiapina, and Neto}{Moura
  et~al\mbox{.}}{2016}]%
        {codivision}
\bibfield{author}{\bibinfo{person}{Fracisco Moura}, \bibinfo{person}{Werney
  Lira}, \bibinfo{person}{Irvayne Ibiapina}, {and} \bibinfo{person}{Pedro
  Neto}.} \bibinfo{year}{2016}\natexlab{}.
\newblock \showarticletitle{Codivision: Uma Ferramenta para Mapear a Divisão
  do Conhecimento entre os Desenvolvedores a partir da Análise de Repositório
  de Código}.
\newblock \bibinfo{journal}{\emph{Congresso Brasileiro de Software - Cbsoft}}
  (\bibinfo{year}{2016}).
\newblock


\bibitem[\protect\citeauthoryear{N.~Zazworka and Seaman}{N.~Zazworka and
  Seaman}{2013}]%
        {zazworka}
\bibfield{author}{\bibinfo{person}{A.~Vetró F.~Shull N.~Zazworka, R.
  O.~Spínola} {and} \bibinfo{person}{C. Seaman}.}
  \bibinfo{year}{2013}\natexlab{}.
\newblock \showarticletitle{A case study on effectively identifying technical
  debt}.
\newblock \bibinfo{journal}{\emph{EASE '13: Proceedings of the 17th
  International Conference on Evaluation and Assessment in Software
  Engineering}} (\bibinfo{year}{2013}).
\newblock


\bibitem[\protect\citeauthoryear{PRESSMAN}{PRESSMAN}{2011}]%
        {PRESSMAN}
\bibfield{author}{\bibinfo{person}{Roger~S. PRESSMAN}.}
  \bibinfo{year}{2011}\natexlab{}.
\newblock \showarticletitle{Software Engineering.}
\newblock \bibinfo{journal}{\emph{New York, NY, USA: McGraw-Hill Science,}}
  (\bibinfo{year}{2011}).
\newblock


\bibitem[\protect\citeauthoryear{S.A.}{S.A.}{2019}]%
        {SonarQube}
\bibfield{author}{\bibinfo{person}{SonarSource S.A.}}
  \bibinfo{year}{2019}\natexlab{}.
\newblock \showarticletitle{SonarQube}.
\newblock \bibinfo{journal}{\emph{Capturado em: http://www.sonarqube.org/,
  abril 2019.}} (\bibinfo{year}{2019}).
\newblock


\bibitem[\protect\citeauthoryear{SPINELLIS}{SPINELLIS}{2005}]%
        {SPINELLIS}
\bibfield{author}{\bibinfo{person}{Diomidis. SPINELLIS}.}
  \bibinfo{year}{2005}\natexlab{}.
\newblock \showarticletitle{Version control systems. Software,}.
\newblock \bibinfo{journal}{\emph{IEEE, v. 22, n. 5,}} (\bibinfo{year}{2005}),
  \bibinfo{pages}{108–--109}.
\newblock


\bibitem[\protect\citeauthoryear{STOREY}{STOREY}{2005}]%
        {STOREY}
\bibfield{author}{\bibinfo{person}{Davor; GERMAN Daniel~M. STOREY,
  Margaret-Anne D.;~ČUBRANIĆ}.} \bibinfo{year}{2005}\natexlab{}.
\newblock \showarticletitle{On the use of visualization to support awareness of
  human activities in software development: a survey and a framework.}
\newblock \bibinfo{journal}{\emph{ACM SYMPOSIUM ON SOFTWARE VISUALIZATION,}}
  (\bibinfo{year}{2005}), \bibinfo{pages}{193–--202}.
\newblock


\bibitem[\protect\citeauthoryear{T.~et al.}{T.~et al.}{2011}]%
        {KLINGER}
\bibfield{author}{\bibinfo{person}{KLINGER T.~et al.}}
  \bibinfo{year}{2011}\natexlab{}.
\newblock \showarticletitle{An enterprise perspective on technical debt}.
\newblock \bibinfo{journal}{\emph{In: ACM. Proceedings of the 2nd Workshop on
  Managing Technical Debt.}} (\bibinfo{year}{2011}).
\newblock


\bibitem[\protect\citeauthoryear{Wohlin and Wesslén}{Wohlin and
  Wesslén}{2000}]%
        {Wohlin}
\bibfield{author}{\bibinfo{person}{Runeson P. Martin Höst M. C. O. Regnell~B.
  Wohlin, C.} {and} \bibinfo{person}{A. Wesslén}.}
  \bibinfo{year}{2000}\natexlab{}.
\newblock \showarticletitle{Experimentation in Software Engineering: An
  Introduction.}
\newblock \bibinfo{journal}{\emph{The Kluwer Internation Series in Software
  Engineering. Kluwer Academic Publishers, Norwell, Massachusets, USA.}}
  (\bibinfo{year}{2000}).
\newblock


\bibitem[\protect\citeauthoryear{ZAKI}{ZAKI}{2003}]%
        {ZAKI:2003}
\bibfield{author}{\bibinfo{person}{Limsoon. ZAKI, Mohammed.~WONG}.}
  \bibinfo{year}{2003}\natexlab{}.
\newblock \showarticletitle{Data Mining Techniques.}
\newblock \bibinfo{journal}{\emph{WSPC/Lecture Notes Series.}}
  (\bibinfo{year}{2003}).
\newblock


\bibitem[\protect\citeauthoryear{Zengyang~Li and Liang.}{Zengyang~Li and
  Liang.}{2015}]%
        {Zengyang:2015}
\bibfield{author}{\bibinfo{person}{Paris~Avgeriou Zengyang~Li} {and}
  \bibinfo{person}{Peng Liang.}} \bibinfo{year}{2015}\natexlab{}.
\newblock \showarticletitle{Software Aging}. In \bibinfo{booktitle}{\emph{A
  systematic mapping study on technical debt and its management.}}
  \emph{(\bibinfo{series}{ICSE '94})}. \bibinfo{publisher}{Journal of Systems
  and Software 101,Supplement C (2015)}, \bibinfo{pages}{193--–220}.
\newblock
\urldef\tempurl%
\url{https://doi.org/10.1016/j.jss.2014.12.027}
\showURL{%
\tempurl}


\end{thebibliography}

\end{document}